\documentstyle[11pt,moriond,epsfig]{article}

\def\NPB{{\em Nucl. Phys.} B}

\def\be{\begin{equation}}
\def\ee{\end{equation}}
\def\bea{\begin{eqnarray}}
\def\eea{\end{eqnarray}}

\begin{document}
\vspace*{4cm}
\title{Next to leading order predictions for $\pi_0\gamma$ and $\pi_0\pi_0$ production at the LHC}

\author{ T. Binoth }

\address{Department of Physics and Astronomy, The University of Edinburgh,\\
Edinburgh EH9 3JZ, Scotland}

\maketitle\abstracts{
The precise knowledge of photon pair production rates
at the LHC is crucial to estimate the sensitivity
to a Higgs boson in the intermediate mass range.
This background consists not only of prompt photons
but also of fake photons stemming from pion decay. 
We present next to leading order predictions for the 
invariant mass distributions of pion photon and pion pairs.}

\section{Introduction}
The results from the LEP experiments suggest that 
the mass of the Higgs boson should lay in a relatively
narrow window. The lower bound from direct searches 
at LEP and the 95\% c.l. indirect upper bound 
is \cite{kawamoto} 
\be\label{bounds}
113.5 \; \mbox{GeV} \; \le M_{Higgs} \le \; 212 \; \mbox{GeV}.
\ee
The upper bound stems from the sensitivity 
of precision observables to quantum corrections
which increase with the Higgs mass. This
means that the discovery of the Standard Model Higgs boson
is to be expected in the near future, though one 
has to say that the validity of the upper bound depends on 
the assumption that the Higgs sector is genuinely
perturbative \footnote{A strongly interacting 
electroweak symmetry breaking sector is not ruled out yet. 
Non-perturbative corrections which are not reliably calculable 
with existing techniques may fake the presence of a light Higgs boson.} 

For the LHC the Higgs mass window (\ref{bounds}) means that 
the decay into photon pairs will be a prominent search channel,
namely in the range 
100 GeV $\le M_{Higgs} \le$ 140 GeV.
If one wants to study the signal significance it is thus
mandatory to know the di--photon background as precisely as possible.

\section{Photon pair production mechanisms at the LHC}
Apart from the Higgs decay into two photons which has a cross
section of around 50 $fb^{-1}$ there are much more important
production mechanism for di--photon events. Although the study of photon
final states is itself interesting I will call them
``background'' throughout this article.   
One can mainly distinguish three classes:
\begin{description}
\item[Direct Photons:] Both photons are produced in a hard interaction.
\item[Photons from fragmentation:] One or both photons are produced 
in the hadronisation of a QCD parton.
\item[Meson decay:] Produced mesons decay into photon pairs which
are misidentified as one single photon in the detector.   
\end{description}
The first two mechanisms are irreducible backgrounds in the sense
that the two photons there can be produced with the same kinematics
as in a Higgs boson decay. Still, the fact that in the second case
at least one photon is accompanied with some amount of hadronic
energy allows to suppress this component to a large extent.
A theoretical next to leading order study for the irreducible 
part of the di--photon background can be found in \cite{BGPW}.
There the Fortran code DIPHOX was used to compute physical
observables using next to leading order matrix elements
for all types of contributions. The same code can be used to
compute the background due to meson decay by replacing 
the photon fragmentation functions by pion fragmentation
functions or other meson fragmentation functions.

The discrimination of photons coming from fragmentation 
and pions is possible because in both cases the electromagnetic
signal will be accompanied with some hadronic energy.
One can now impose isolation criteria by 
not allowing more than a certain amount
of hadronic energy, $E_{T max}$, in a cone, 
$R=\sqrt{(\Delta\phi)^2+(\Delta y)^2}$ defined in
rapidity and azimuthal space around the photon/meson direction.

\section{Next to leading order predictions for $\pi_0\gamma$ and 
$\pi_0\pi_0$}
We now present next to leading order predictions for pion pair
and pion photon production at the LHC. The 
remaining uncertainties for these predictions are
three--fold. First there are theoretical  uncertainties due to missing
higher order corrections. They are typically estimated
by varying unphysical scales present in the calculation around
some value chosen for the hard scale. In our case
one has a dependence on the renormalisation scale, $\mu$, the
factorisation scale, $M$, and the fragmentation scale, $M_f$.
A detailed analysis of scale dependencies will be presented 
elsewhere \cite{BGPW_pion}.   
A second uncertainty comes from the fragmentation functions
used. Typically they are parametrised by data samples
in the intermediate z range where z is the energy fraction of meson 
to parent parton. A recent parametrisation  \cite{kkp}
was used for the plots below. 
Imposing severe isolation criteria to suppress
background photons and mesons from fragmentation  imposes
a lower bound on this variable, namely
\be
z > z_{min} = \frac{p_{T min}}{p_{T min} + E_{T max}}.
\ee 
Here $p_{T min}$ is the lower experimental cut--off for 
the transverse momentum of the observed boson, typically
25 GeV at the LHC. Obviously severe isolation criteria as
e.g. $E_{T max}=5$ GeV in a cone $R = 0.4$ 
restrict the z range to values near 1
where the fragmentation functions are poorly known. This induces
a third uncertainty on the theoretical side. 
One has to be aware that potentially 
large logarithms $\sim\log(1-z)$ enter the game 
which may have to be resummed to give a reliable prediction.
All these issues deserve further investigation \cite{BGPW_pion}. 

\begin{figure}[ht]
\hspace{2cm}
    \epsfxsize = 12.5cm
    \epsffile{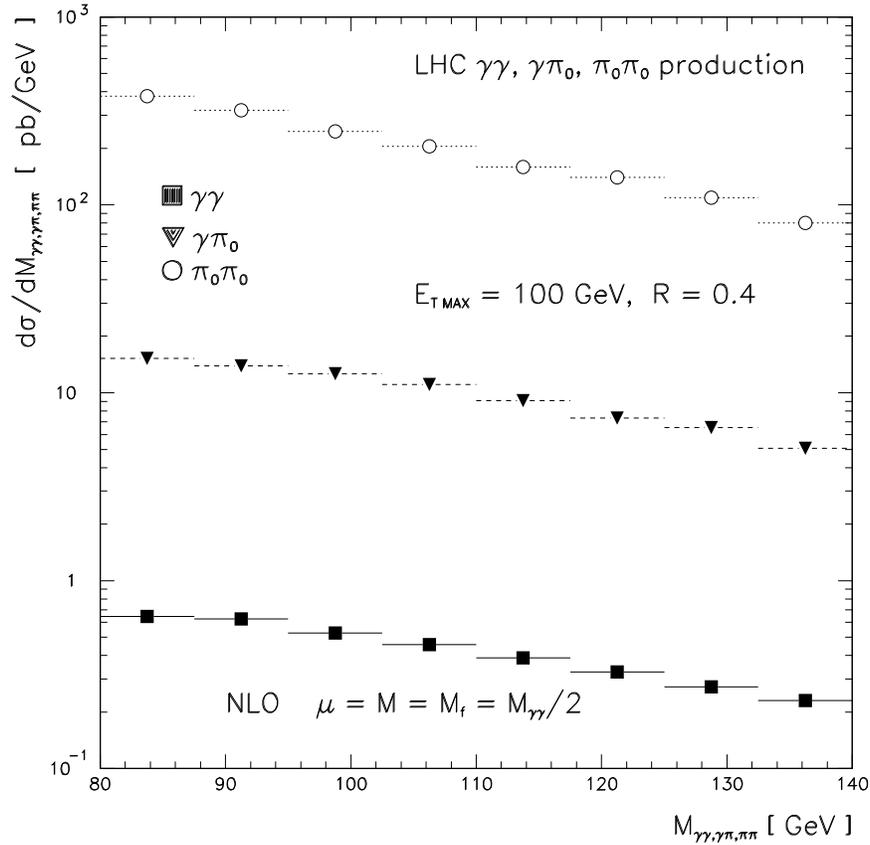}
\caption{\label{Fig:all} Production rates of 
$\gamma\gamma$, $\gamma\pi_0$, and $\pi_0\pi_0$ at NLO
with isolation cut: $E_{T max} = 100$ GeV in cone $R = 0.4$.} 
\end{figure}

In the plots presented below
we have chosen $\mu=M=M_f=M_{bb}/2$, where $M_{bb}$ is the 
invariant mass of the produced boson pair, $b=\gamma,\pi_0$. 
We used the MRST2 \cite{MRST} structure functions. 
Fig. \ref{Fig:all} shows the invariant mass distribution 
of $\gamma\gamma$, $\gamma\pi_0$, $\pi_0\pi_0$ for
a very loose isolation criteria, $E_{T max}=100$ GeV in a cone
of $R=0.4$. In addition to the isolation cuts,
standard rapidity and $p_T$ cuts were applied:
$| y(boson_{1,2}) | < 2.5 $, $p_T(boson_1)>40$ GeV, and
$p_T(boson_2)>25$ GeV.
Fig. \ref{Fig:all} gives an idea of the large suppression factors
which are needed for the reducible backgrounds 
to find a Higgs signal in the given mass window.
Though being a reducible background the huge 
production rates for pions make a reliable understanding
of these rates necessary.

\begin{figure}[ht]
\hspace{2cm}
    \epsfxsize = 13cm
    \epsffile{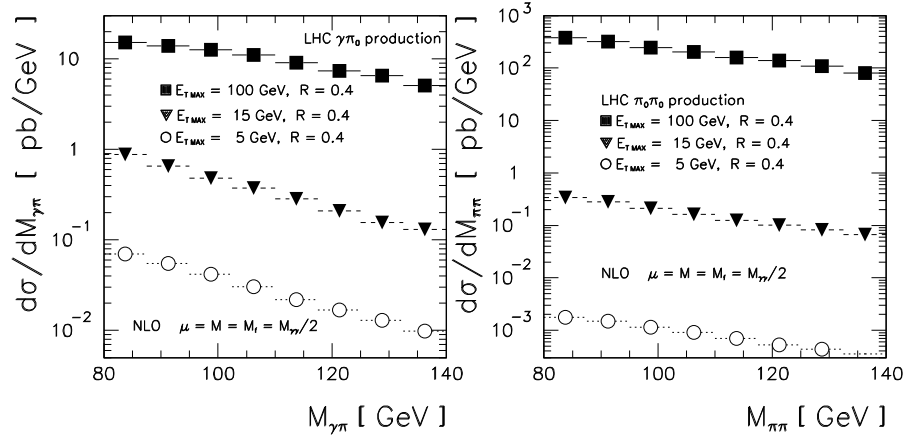}
\caption{\label{Fig:isol} Effect of isolation on the 
$\gamma\pi_0$ (left), $\pi_0\pi_0$ (right) invariant mass distribution.}
\end{figure}

In Fig. (\ref{Fig:isol}) the dependence on isolation criteria for 
$\pi_0\gamma$ (left) and $\pi_0\pi_0$ production (right)   
is shown. Again, in addition to the isolation cuts, standard cuts
on the observed bosons as defined above are applied. 
With an isolation criteria of $E_{T max} = 15$ GeV in a cone $R=0.4$
the rates for $\gamma\gamma$, $\gamma\pi_0$, $\pi_0\pi_0$ 
are all still of order pb/GeV. Increasing the cut (lowering
$E_{T max}$) does not change the $\gamma\gamma$
background much from now on, as dominantly direct 
photons are present already which are insensitive to the cut. 
On the other hand the pion rates are further reduced. 
Whereas the differential cross section $d\sigma/dM_{\pi\pi}$ 
for $\pi_0\pi_0$ is of the order
fb/GeV for a isolation criteria of $E_{T max} = 5$ GeV, the
differential cross section for $d\sigma/dM_{\pi\gamma}$
is still 1 to 2 orders of magnitude higher and because of
the uncertainties mentioned above still a dangerous
background. One should always bare in mind that
the Higgs signal will only be around 50 $fb^{-1}/GeV$,
if one  assumes that the signal is located
inside a 1 GeV bin. 
It should be said that experimentally
further reduction factors for fake photons coming from
meson decays can be applied. These are typically 
of order 2 for any 
misidentified pion \cite{KatisPhdThesis}.     
We note that the next to leading order corrections
are typically 50\%--100\% of the leading order
contribution but general quantitative statements
of their sizes may be dangerous in connection with 
isolation criteria. This will be discussed elsewhere \cite{BGPW_pion}.    
  
\section{Conclusion}
Next to leading order results for pion photon and pion pion
production at LHC are presented. The present calculation pins
down theoretical uncertainties for the production
rates of these processes. This is important, as they are
huge backgrounds for Higgs searches in the di--photon channel.
Isolation  criteria  suppress pion pair production 
considerably but the pion photon channel remains still
relevant. We remarked that for stringent isolation 
cuts the knowledge of the pion fragmentation functions
at the high z end is mandatory and also that --- on
the theoretical side --- resummation of $\log(1-z)$ terms
could become important.

\section*{Acknowledgments}
I would thank my collaborators J.~Ph.~Guillet, E.~Pilon, and
M.~Werlen for giving me the opportunity to present our results at 
the Moriond conference.\\
This work was supported in part by the EU Fourth Training Programme
"Training and Mobility of Researchers", Network "Quantum Chromodynamics
and the Deep Structure of Elementary Particles", 
contract FMRX--CT 98--0194 (DG 12 -- MIHT).

\end{document}